\documentclass[12pt]{article}
\usepackage{epsfig}
\usepackage{epsfig}
\begin{document}
\def\tr{{\rm tr}\, }
\def\Tr{{\rm Tr}\, }
\def\hTr{\hat{\rm T}{\rm r}\, }
\def\be{\begin{eqnarray}}
\def\ee{\end{eqnarray}}
\def\ctt{\chi_{\tau\tau}}
\def\cta{\chi_{\tau a}}
\def\ctb{\chi_{\tau b}}
\def\cab{\chi_{ab}}
\def\cba{\chi_{ba}}
\def\ptt{\phi_{\tau\tau}}
\def\pta{\phi_{\tau a}}
\def\ptb{\phi_{\tau b}}
\def\>{\rangle}
\def\<{\langle}
\def\d{\hbox{d}}
\def\pab{\phi_{ab}}
\def\lb{\label}
\def\appendix{{\newpage\section*{Appendix}}\let\appendix\section%
        {\setcounter{section}{0}
        \gdef\thesection{\Alph{section}}}\section}
\renewcommand{\figurename}{Fig.}
\renewcommand\theequation{\thesection.\arabic{equation}}
\hfill{\tt }\\\mbox{}
\vskip0.3truecm
\begin{center}
\vskip 2truecm {\Large\bf Entanglement entropy and the Ricci flow}
\vskip 1.5truecm
{\large\bf
Sergey N.~Solodukhin\footnote{
{\tt s.solodukhin@iu-bremen.de}}
}\\
\vskip 0.6truecm
\it{School of Engineering and Science, \\
International University Bremen, \\
P.O. Box 750561,
Bremen 28759,
Germany}
\end{center}
\vskip 1cm
\begin{abstract}
\noindent We analyze the Ricci flow of a noncompact
metric that describes a two-dimensional black hole. We consider
entanglement entropy of a 2d black hole which is due to
the quantum correlations between two subsystems: one is inside and
the other is outside  the black hole horizon. It is demonstrated
that the entanglement entropy is monotonic along the Ricci flow.
\end{abstract}
\vskip 1cm
\newpage

\section{Introduction}
\setcounter{equation}0
The  irreversibility of the observable phenomena
has always  been an intriguing fact that needed an explanation.
In quantum field theory the irreversibility is usually associated with
the Renormalization Group (RG) flow.  Although it is  very difficult
to give a general proof of the irreversibility this proof exists in two
space-time dimensions and is known as the c-theorem
\cite{Zamolodchikov:1986gt}. For 2d $\sigma$-models, as those arising in
string theory, this proof assumes compactness of the target space \cite{Tseytlin:1987bz}, \cite{Polchinski:1987dy}.
Generalization of the c-theorem to higher dimensions and more general class of theories
is not straightforward. In general,  a rather complicated and even
chaotic behavior of the RG flow is possible
\cite{Morozov:2003ik}.

The irreversibility of time evolution is usually
associated with the notion of entropy  in a coarse-grained
description of system when the detail knowledge about the individual states
is lost for the advantage of having an effective description of a
huge number of such states. Yet the entropy can be associated with geometry
in the case the coarse-grained states are residing entirely inside a compact surface.
This entropy is known as entanglement (or geometric) entropy and is due to the
quantum correlations between the subsystems divided by the surface.
Entanglement entropy is determined by geometry of the surface and, in this
sense, reminds the Bekenstein-Hawking entropy of black holes. In fact, the study
of entanglement entropy in the past was mainly motivated by the hope to find a
statistical-mechanical explanation to the black hole entropy
\cite{Bombelli:1986rw}-\cite{Frolov:1996hd}. Recently, entanglement entropy
has been in the focus of studies in condensed matter as an appropriate quantity that
measures   quantum entanglement in the quantum-mechanical systems
\cite{Korepin}-\cite{Latorre:2004pk}. Although a great deal is known about
entanglement entropy of various systems its irreversibility was not
fully revealed. A natural guess is that the irreversibility should be
understood with respect to the RG flow. A relation between
entanglement entropy in two-dimensional field theory and the c-theorem was
discussed in \cite{Casini:2004bw}. On the other hand,
the recent work \cite{Latorre:2004pk}  investigates how
entanglement entropy of certain two-dimensional spin chain models
changes under the RG flow. It is found that the entanglement entropy is
monotonically decreasing  along the RG trajectory in this case. This is
interpreted as loss of entanglement in the infra-red regime. The RG flow is not directly relevant to
entanglement entropy of black holes since the entropy is entirely due to
the presence of the black hole and is  nontrivial even in the
case when interactions, and, hence, the non-trivial renormalization group flow
are absent in the field system.

In this note we suggest that entanglement
entropy of a black hole decreases monotonically under the Ricci flow. The latter is analogous
to the RG flow. These two flows are, in fact, related for the two-dimensional
$\sigma$-models with the Riemann manifold as  target space
\cite{Friedan:1980jf}. The Ricci flow \cite{Hamilton}, \cite{Hamilton2}
plays an important role in  the problem of geometrization of 3-dimensional
manifolds. The recent enthusiasm regarding the Ricci flow  has been sparked of course
by works of Perelman \cite{Perelman}. For elementary review on the Ricci flow
see \cite{BChow}, \cite{BChow2}, a review on the recent development is
\cite{Anderson}.
Some applications of the Ricci flow and, in
particular, of Perelman's results to the physics models are discussed in
\cite{Bakas:2004hn}-\cite{Headrick:2006ti}. Our main focus in this note is on the
two-dimensional case when the metric evolving under the Ricci flow
describes a 2d black hole. We first remind the reader some  facts about
entanglement entropy and present an exact expression for the entropy of 2d
black hole which is due to  a two-dimensional conformal field. We then discuss the
properties of the Ricci flow in two dimensions putting focus on whether the
property of metric to describe black hole changes under the flow.
We finally show that entanglement entropy is monotonically decreasing along the
Ricci flow provided the scalar curvature of the initial metric is positive.
Our result looks similar to the one obtained in \cite{Latorre:2004pk} for the
RG flow and, perhaps, there is a  reason why the entanglement entropy should
decrease.

\section{Entanglement entropy of 2d black holes}
\setcounter{equation}0 Entanglement entropy is defined for a
system divided into two  subsystems (L and R) and is due to the
short-range correlations that are present in the system. The notion of
entanglement entropy is ideally suited for black holes since the
black hole horizon naturally divides the space-time on two
subsystems so that an observer outside a black hole  does not have
access to excitations propagating inside the horizon. Assume, that
the whole system of a quantum field $\phi$, that takes values
$\phi_L$ and $\phi_R$ in the subsystems L and R respectively, is
in a pure ground state
\begin{equation}
\Psi_0=\Psi_0 (\phi_R,\phi_L) \label{22.1}
\end{equation}
that is  the functional of both $\phi_R$ and $\phi_L$ modes. For
an observer  who has access only to one subsystem, say R, it is
more reasonable to introduce  a density matrix
\begin{equation}
\rho (\phi_R^1, \phi_R^2)=\int{} [ {\cal D} \phi_L] \Psi^*_0
(\phi_R^1, \phi_L) \Psi_0 (\phi^2_R, \phi_L) \label{22.2}
\end{equation}
where one traces over all modes $\phi_L$. Entanglement entropy is
defined as
\begin{equation}
S_{}=-\Tr  \hat{\rho}  \ln \hat{\rho}=(-\alpha \partial_\alpha
+1) \Tr {\rho}^\alpha |_{\alpha=1}, \   \ \hat{\rho}={\rho
\over \Tr \rho}~~. \label{22.3}
\end{equation}

Applying this construction to a 2d black hole (for a general case
see \cite{Solodukhin:1995ak}),  we identify  all  modes inside the
black hole horizon with L-modes that are to be traced over. The
ground state of the black hole is given by  the Euclidean
functional integral  \cite{BFZ} over fields defined on a half of
the Euclidean black hole space-time
\begin{equation}
ds^2_{E'}=\beta^2_H f(x) d\varphi^2 +{1\over f(x)}dx^2 ~~,
\label{2.4}
\end{equation}
defined for values   $-{\pi \over 2} \leq \varphi \leq {\pi \over
2}$ of angular coordinate $\varphi$.  The inverse Hawking
temperature $\beta_H=T^{-1}_H$ is determined by the derivative of
the metric  function $f(x)$ on the horizon ($f(x_+)=0$,
$\beta_H={4\pi \over f'(x_+}$). Functions $\phi_R$ and $\phi_L$
that enter as arguments in (\ref{22.1}) are the fixed values at
the boundaries $\phi_R=\phi (\varphi={\pi \over 2})$; $\phi_L=\phi
(\varphi=-{\pi \over 2})$, giving the boundary condition in the
path integral. The density matrix $\rho (\phi^1_R, \phi^2_R)$,
obtained by tracing over $\phi_L$-modes, is defined by the path
integral over fields on the full black hole instanton $E, \  (-{3
\over 2}\pi\leq\varphi\leq {1 \over 2}\pi)$, with a cut along the
$\varphi={\pi \over 2}$ axis and taking values $\phi^{1,2}_R$
above and below the cut. The trace $\Tr \rho$ is obtained by
equating the fields across the cut and doing the unrestricted
Euclidean path integral on the complete black hole instanton $E$.
Analogously, $\Tr \rho^n$ is given by the path integral over
fields defined on $E_n$, a $n$-fold cover of $E$. Thus,  $E_n$ is
the manifold with an abelian isometry (generated by vector
${\partial }_\varphi$) with horizon  $\Sigma$ as a stationary
point. Near $\Sigma$ the $E_n$ looks as a cone $E_n= {\cal C}_n$
with tip at $\Sigma$ and  the angle deficit $\delta=2\pi(1-n)$.
This construction can be analytically continued to arbitrary
(non-integer) $n \rightarrow \alpha={\beta \over \beta_H}$.

The calculation of entanglement entropy thus reduces to a
calculation of the functional integral on a gravitational
background with a conical singularity. In two dimensions, for a
conformal field, the result of the functional integration is the
non-local  Polyakov action. The entropy calculation thus can be
carried out explicitly. For a black hole metric written in the
conformal form $g_{\mu\nu}=e^{2\sigma}\delta_{\mu\nu}$
 entanglement entropy takes the form \cite{RCM}, \cite{Fursaev:1995ef}
$S={c\over 6}\sigma (x_+)+{c\over 6} \ln (\Lambda/ \epsilon )$,
where $x_+$ is the location of the horizon and $\epsilon$ is an UV
regulator. The complete black hole geometry can be described by a 2d metric
in the Schwarzschild like form
\begin{equation}
ds^2_{bh}=f(x)d\tau^2+{1\over f(x)}dx^2~~,
\label{B}
\end{equation}
where the metric function $f(x)$ has a simple zero in $x=x_+$;  $x_+\leq x \leq L$,
 $0\leq \tau
\leq \beta_H$, $\beta_H={4\pi \over f'(x_+)}$.
It is easy to see that (\ref{B}) is conformal to the flat disk of radius $z_0$
($\ln z={{2\pi \over \beta_H}\int^x_L {dx \over f(x)}}$):
\begin{eqnarray}
&&ds^2_{bh}=e^{2\sigma }z_0^2(dz^2+z^2d\tilde{\tau}^2 )~~, \\
&&\sigma={1\over 2} \ln f(x)-{2\pi \over \beta_H}\int^x_L {dx\over f(x)}
 +\ln {\beta_H  \over 2\pi z_0}~~, \nonumber
\label{4}
\end{eqnarray}
where
 $\tilde{\tau}={2\pi \tau \over \beta_H}$ ($0\leq \tilde{\tau} \leq2\pi$),
 $0 \leq z\leq 1$.
So that entanglement entropy of a 2d black hole takes the form
 \cite{Frolov:1996hd} (see also \cite{Solodukhin:1994yz}
and  \cite{Solodukhin:1995te})
\begin{equation}
S={c\over 12} \int^L_{x_+}{dx \over f(x)}({4\pi \over \beta_H}-f')+
{c\over 6}\ln ({\beta_H  f^{1/2}(L)\over \epsilon}) ~~,
\label{S}
\end{equation}
where we omitted the irrelevant term which is function of $(\Lambda , z_0)$
but not of the parameters of the black hole and have retained dependence on UV
regulator $\epsilon$.

In the above analysis it is assumed that the black hole resides
inside a finite size box  and $L$ is the value of $x$-coordinate of the
boundary of the box. The coordinate invariant size of the
subsystem outside the black hole is $L_{\tt
inv}=\int_{x_+}^Ldx/\sqrt{f(x)}$. Entanglement entropy is a
monotonic function of  the size of the box,
\be
\partial_{L_{\tt inv}}S={c\over 12}{4\pi\over \beta_H}{1\over \sqrt{f(L)}}>0~~.
\lb{monS}
\ee
For small $L_{\tt inv}$ one  finds an
expansion \cite{Solodukhin:2006xv}
\be
S={c\over 6}\left(\ln{L_{\tt inv}\over \epsilon}-{R(x_+)\over 36}L^2_{\tt
    inv}\right)~~,
\lb{Lsm} \ee where $R(x_+)$ is the value of the Ricci scalar at
the horizon. If the black hole space-time is asymptotically flat
then entanglement entropy approaches the entropy of the
two-dimensional thermal gas, $S_{\tt th}={c\pi\over 3}L_{\tt
inv}T_H$, in the limit of  large $L_{\tt inv}$. Thus, entropy
(\ref{S}), in spite the fact that it is essentially due to the
presence of the black hole horizon,  contains an extensive
component that carries  information about the global structure of
the black hole geometry.

\section{The Ricci flow in two dimensions}
\setcounter{equation}0

\noindent {\bf The Ricci flow of static metric}. In two dimensions the Ricci
flow takes the form
\be
\partial_\lambda g_{ij}=-R g_{ij}~~,
\lb{2.1}
\ee
where $R$ is the Ricci scalar of metric $g_{ij}$. We
choose  that initial (at $\lambda=0$)
metric  describes a static black hole  in the Schwarzschild like
 form (\ref{B})
\be
ds^2=g(x) d\tau^2+{1\over g(x)}dx^2~~. \lb{2.2}
\ee
The metric
function $g(x)$ is positive everywhere except the horizon where it
becomes zero. If black hole is non-degenerate than near this
point, say $x=x_+$, one has that $g(x)=b_0(x-x_+)+O((x-x_+)^2)$ so
that the Hawking temperature $T_H=b_0/
  4\pi$ is non-vanishing.  Metric (\ref{2.2}) has a Killing vector $\partial_\tau$ that generates the shifts along the
time direction. In the Euclidean case the $\tau$ coordinate should
be identified with period $1/T_H$ so that the Killing vector
generates rotations. Asymptotically flat ($g\rightarrow 1$ at
infinity) metric (\ref{2.2}) is topologically  disk.
Geometrically it looks like a cigar with a tip (where metric is
completely regular) at the  horizon.

For $\lambda>0$ the rotational symmetry is preserved so that the
appropriate form for the metric is
\be ds^2=g(x,\lambda)
d\tau^2+{e^{2\phi (x,\lambda)}\over g(x,\lambda)}dx^2~~, \lb{2.3}
\ee
where $g(x,\lambda)$ and $\phi (x,\lambda)$ are now functions
of both $x$ and $\lambda$. The "initial" value for $\phi$ is
$\phi(x,\lambda=0)=0$. The Ricci scalar for metric (\ref{2.3})
reads \be R=e^{-2\phi(x)}(-g''_x+\phi'_xg'_x)~~. \lb{R} \ee The
Ricci flow (\ref{2.1}) for metric (\ref{2.3}) reduces to a set of
equations
\be
&&\partial_\lambda g=-ge^{-2\phi}(-g''_x+\phi'_xg'_x)~~, \nonumber \\
&&\partial_\lambda e^{2\phi}=2(g''_x-\phi'_xg'_x)~~. \lb{eqs} \ee
Combining these two equations one finds that \be {\partial_\lambda
g\over g}=\partial_\lambda \phi~~. \ee Taking into account the
``initial'' condition for $\phi(x,\lambda)$ one has \be
e^{\phi(x,\lambda)}={g(x,\lambda)\over g(x)}~~. \lb{phii}
\ee
 So that the metric
(\ref{2.3})  takes the form
\be
ds^2=g(x,\lambda)\left(d\tau^2+{1\over g^2(x)}dx^2\right)~~.
\lb{metric} \ee
The function $g(x,\lambda)$ satisfies equation
\be
\partial_\lambda g=g_0\left({g_0g'_x\over g}\right)'_x~~,
\lb{eq2} \ee
where we denoted $g_0(x)\equiv g(x)$.

\bigskip

\noindent {\bf Transformation to the Schwarzschild like form. The
Ricci-DeTurck flow.} For $\lambda \neq 0$ metric (\ref{metric})
does not take the Schwarzschild like form (\ref{B}) anymore. But
it can be brought to this form by a $\lambda$-dependent coordinate
transformation. Introduce a new coordinate $y$,
\be
y=\int_{x_+}^x{g(x,\lambda)\over g_0(x)}dx~,~~\partial_\lambda
y=(g'_y-b_0)~~.
\lb{y}
\ee
We choose limits of integration in
(\ref{y}) so that $y=0$ corresponds to $x=x_+$, the location of
the horizon in the initial metric (\ref{2.2}). In terms of the
coordinate $y$  metric (\ref{metric})  takes the Schwarzschild
like form (\ref{B})
\be
ds^2=g(y,\lambda)d\tau^2+{1\over
g(y,\lambda)}dy^2~~.
\lb{metric2}
\ee
The flow equation for
$g(y,\lambda)$  reads \be
\partial_\lambda g=gg''_y-g'^2_y+b_0g'_y~~,
\lb{eq3}
\ee
where $b_0=g'_0(x)|_{x=x_+}$.

Since we applied a $\lambda$-dependent coordinate transformation (\ref{y})
when derived metric (\ref{metric2}) the components of the transformed metric
satisfy a modified flow equation (\ref{eq3}) known as the Ricci-DeTurck flow,
\be
\partial_\lambda g_{ij}=-Rg_{ij}-\nabla_i\xi_j-\nabla_j\xi_i~~,
\lb{eq4}
\ee where
$\xi_i$ is the deformation vector. In the case
at hand we have
\be \xi_\tau =0~,~~\xi_y={1\over
g(y,\lambda)}(g'_y-b_0)~~. \lb{xi} \ee Note that this is a
gradient vector, \be \xi_t=-\partial_\tau\psi(y,\lambda)~,~~
\xi_y=-\partial_y
\psi(y,\lambda)~,~~\psi(y,\lambda)=\int^y{d\tilde{y}\over
g(\tilde{y},\lambda)}(b_0-g'_y)~~. \lb{grad}
\ee
In fact, $\psi$
is a rotation invariant solution to equation
\cite{Solodukhin:1994yz}, \cite{Solodukhin:1995te}
\be \Delta
\psi=R \lb{psi} ~,~~ R=-g''_y
\ee and is a non-local object,
$\psi={1\over \Delta}R$. Choosing the integration constant in
(\ref{grad}) appropriately, one can express entanglement entropy
(\ref{S})  in terms of $\psi(y)$ \cite{Solodukhin:1995te}, \be
S=-{c\over 12}\psi(0))+{c\over 6}\ln(1/\epsilon)~~. \lb{Sagain}
\ee
For metric (\ref{metric}) we have that $\psi(x,\lambda)=\ln{g(x)\over
  g(x,\lambda)}+\psi_0(x)$, where $\psi_0(x)$ is a solution to equation
(\ref{psi}) for metric $g_0(x)$.

\bigskip

\noindent {\bf The near-horizon analysis.} Assume that near $y=0$
the function $g(y,\lambda)$ can be presented in the form of the
Taylor series \be g(y,\lambda)=\sum_{n=0}^\infty a_n(\lambda)
y^n~~, \lb{series} \ee where we included the constant term with
$n=0$. This term is absent in the initial metric, $a_0=0$ at
$\lambda=0$, but may appear for some $\lambda>0$. If it appears
then this means that the point $y=0$ is not a horizon anymore. As
we show below, this does not happen and the point $y=0$ stays to
be horizon for all $\lambda>0$.

Substituting expansion (\ref{series}) into equation (\ref{eq3}) we
get the flow equations for the coefficients in the Taylor
series
\be
&&n=0~: \ \ \ \partial_\lambda a_0=(b_0 -a_1)a_1+2a_0 a_2  \lb{sereq}\\
&& n=1~: \ \ \ \partial_\lambda a_1=2(b_0-a_1)a_2+6a_3 a_0 \nonumber \\
&& n>1~:\ \ \ \partial_\lambda
a_n=(n+2)(n+1)a_0a_{n+2}+(n+1)(b_0-a_1)a_{n+1}\nonumber \\
&&\ \ \ \ \ \ \ \ \  \ \ \ \ \ +(n^2-1)a_1
a_{n+1}+\sum_{m=2}^{n}m(2m-n-3)a_m a_{n-m+2}\nonumber  \ee At
$\lambda=0$ we have that
 $a_0=0$ and $a_1=b_0$. Substituting this
into the  first two equations in (\ref{sereq})  and after a number
of differentiations we find that $\partial^{(k)}_\lambda a_0=0$
and $\partial_\lambda^{(k)} a_1=0$, $k\geq 1$ at $\lambda=0$. From
this we conclude that $a_0=0$ and $a_1=b_0$ for all $\lambda$
(expanding $a_0$ and $a_1$ in Taylor series in $\lambda$ one finds
that  all terms, except first, in the series vanish). The value of
$y$-derivative of metric function in (\ref{metric2}) at $y=0$
gives the Hawking temperature $T_H={1\over 4\pi}g'_y(y=0)$ of the
horizon. The fact that this value does not change means that the
Hawking temperature remains constant under the Ricci flow,
$T_H(\lambda)=T_H(\lambda=0)$.

\bigskip

\noindent {\bf The maximum principles.} The Ricci scalar $R$ is an
important quantity to look at since its evolution preserves certain bounds.
Under the Ricci-DeTurck flow
(\ref{eq4}) the Ricci scalar changes according to equation \be
\partial_\lambda R=-\xi^k\partial_k R+ \Delta R+R^2~~.
\lb{RR}
\ee
By applying the maximum principle to this equation one obtains important bounds on the curvature of the
space-time under the flow.
In terms of metric (\ref{metric2}) and vector field (\ref{xi})
equation (\ref{RR}) takes the form
\be
\partial_\lambda R=b_0 R'_y+gR''_y +R^2~~.
\lb{RR1} \ee
At a point in which $R(y,\lambda)$ takes a local
minimum one has $R'_y=0$ and $R''_y>0$ and hence $\partial_\lambda
R>0$. The minimal value of scalar curvature is thus increasing
with $\lambda$. Therefore, if the initial metric (at $\lambda=0$) has
everywhere positive curvature $R(y)>0$, $y>0$ then
for any $\lambda>0$  we have that $R_{\tt min}>0$. It does not immediately
follow that the the Ricci scalar is everywhere positive for $\lambda>0$
since equation (\ref{RR1}) is defined on a half-line and hence one
should check that values of $R$ at the boundaries do not become negative.
We assume that metric is asymptotically flat so that $R=0$ at $y=+\infty$.
On the other boundary,
at $y=0$, where $g(y)$ vanishes, one finds that
\be
\partial_\lambda R(0)=b_0R'_y(0)+R^2(0)~~.
\lb{R0}
\ee
There are two cases to consider. 1) $R'_y(0,\lambda )<0$, then
$R(0,\lambda)>R_{\tt min}>0$ if there is a local minimum at some point $y>0$
or $R(0,y)>R(y=\infty)=0$ if no such point exists.
2) $R'_y(0)>0$, then $R(0,\lambda)$ is increasing with
$\lambda$ and hence $R(0,\lambda)>0$ as well. Thus, the value of $R$ at $y=0$ remains
positive in any case.
The property of the Ricci flow to preserve the positive sign of curvature is
important for our discussion in the next section.

The maximum principle is an useful tool. It can be used  to prove that there can be
no more horizons than in the initial metric. Suppose that the
initial metric function $g_0(y)$  has a horizon at $y=0$  and is
asymptotically flat. So it
changes from zero at $y=0$ to 1 at $y=+\infty$ and is positive everywhere in between. Can there, for
some $\lambda>0$, appear another point $y=y_1>0$ at which
$g(y_1,\lambda)=0$? If there is such a point it forms from a
minimum of the function $g(y,\lambda)$. At a local minimum one has
that $g'_y=0$ and $g''_y>0$. Then from equation (\ref{eq3}) one
finds that $\partial_\lambda g=g g''_y>0$ and hence the value of
function $g(y,\lambda)$ at the minimum is increasing. This means
that it never can reach zero so that no new horizon will be
formed.

On the other hand, if $g(y,\lambda)$ has a local maximum then
$g'_y=0$ and $g''_y<0$ at that point so that $\partial_\lambda g=g
g''_y<0$. Thus, the maximal value of the function $g(y,\lambda)$ is
decreasing with $\lambda$. The Ricci flow thus tends to smear out
the metric function $g(y,\lambda)$ over the half line $y>0$.

The tendency of the Ricci flow to smooth out the initial geometry
works in an interesting way. We could have started (at $\lambda=0$) with a black hole
at a temperature $\beta^{-1}$ different from the Hawking temperature,
$\beta\neq \beta_H$. The initial geometry then would have a conical singularity at
the horizon. As the example of the ``decaying cone'' solution found in \cite{Adams:2001sv},
\cite{Gutperle:2002ki}
shows, the geometry evolving under the Ricci flow then would be completely
regular for any $\lambda>0$ so that the temperature is immediately
switched to the Hawking temperature under the flow.

\bigskip

\noindent {\bf Asymptotically flat metrics.} At infinity of
asymptotically flat space-time one has $g=1+h$, $h\ll 1$.  $h$
satisfies the  heat type equation
\be
\partial_\lambda h=h''_y+b_0h'_y~~.
\lb{h} \ee
It is  a linear equation that is also satisfied for the
derivatives of function $h(y,\lambda)$. Suppose that the initial data for $h$
falls off by a power law, $h_0(y)\sim {1\over y^k}$,
$k \geq 1$  for large $y$. Then the term $h''_y$ in (\ref{h}) can be neglected  as small
 so we have that $h(y,\lambda)$ satisfies equation \be
\partial_\lambda h=b_0h'_y~~.
\lb{h1} \ee Taking into account the initial condition we get \be
h(y,\lambda )=h_0(y+b_0\lambda)\sim {1\over (y+b_0\lambda)^k }~~
\lb{sol} \ee for the solution. Clearly, it decays both for large
$y$ and large $\lambda>0$. For $\lambda>0$ the bound
$|h(y,\lambda)|\leq |h_0(y)|$ is satisfied. Same is true for the
scalar curvature $R=-h''_y(y,\lambda)$.
Thus, the initially asymptotically flat metric remains to be
asymptotically flat  under the Ricci flow.

\bigskip

\noindent {\bf The stationary point (cigar soliton).} Summarizing
our analysis   so far we have found that

\noindent 1. the black hole metric at $\lambda=0$ remains to be a black hole for $\lambda>0$;

\noindent 2. the Hawking temperature $T_H$ does not change under the Ricci flow;

\noindent 3. asymptotically flat metric at $\lambda=0$ remains
to be asymptotically flat
   for $\lambda>0$;

\noindent 4. no new horizons are formed.

We can draw an  important conclusion from these observations. The
end point (if there is one) of the Ricci flow that started with an
asymptotically flat 2d black hole metric is a non-constant curvature
space-time that has same temperature as the initial geometry and
is asymptotically flat.  This is different from how the Ricci flow
behaves in the case of compact manifold. In the latter case the
Ricci flow ends on a geometry with constant (positive or negative)
scalar curvature \cite{Hamilton2}.

The stationary point of  equation (\ref{eq3}) is easy to find.
The right hand side of equations (\ref{eq3}) vanishes identically
for a metric function
\be g_{\tt st}(y)={b_0\over
\Lambda}(1-e^{-\Lambda y})~~, \lb{gst}
\ee where $\Lambda$ is
integration constant. Demanding $g_{\tt st}=1$ for $y=+\infty$ we have that
$\Lambda=b_0$. This metric is the so-called Ricci soliton. It was
found in \cite{Hamilton2}. In physics this metric is known as 2d
black hole that appears in the context of two-dimensional string
theory \cite{Witten:1991yr}, \cite{Mandal:1991tz} . It solves the
one-loop renormalization group equations in certain $\sigma$-model
with non-compact target space. The function
$\psi=\Lambda y$ (\ref{psi})  is related to the linear
dilaton.  Note that this metric has the everywhere positive
curvature and is sometimes called the cigar soliton. It is asymptotic
to a flat cylinder at infinity and has maximal curvature at the
origin. Depending on the initial metric the flow may or may not
approach the stationary point. An example of the flow that
approaches the stationary point is given in \cite{Hori:2001ax}.

It should be noted that (\ref{gst}) is a stationary point of the Ricci-DeTruck
flow (\ref{eq4}) but not of the original Ricci flow equation (\ref{2.1}), (\ref{eq2}).
Indeed, solving the equation (\ref{y}) we find a relation between coordinates
$y$ and $x$, so that at $\lambda=0$ we have that $y=x$. The corresponding solution to
equation (\ref{eq2})  is given by function $g(x,\lambda)=g_{\tt
  st}(y(x,\lambda))$,
\be
g(x,\lambda)={1-e^{-b_0x}\over 1+(e^{b_0^2\lambda}-1)e^{-b_0x}}~~.
\lb{gx}
\ee
The metric (\ref{metric})
\be
ds^2={1\over
  1+(e^{b_0^2\lambda}-1)e^{-b_0x}}\left((1-e^{-b_0x})d\tau^2+{dx^2\over (1-e^{-b_0x})}\right)
\lb{ggg} \ee then  solves the Ricci flow equation (\ref{2.1}) and
is a conformal, $\lambda$-dependent, deformation of the stationary
metric (\ref{metric2}), (\ref{gst}). This metric is eternal
solution to the Ricci flow equations. It extrapolates between flat
metric at $\lambda=-\infty$ and a constant curvature ($R=b_0^2$)
metric at $\lambda=+\infty$. In these limits one focuses on
various regions of the cigar soliton: asymptotic infinity or the
near horizon (tip of cigar) region.

\bigskip

\noindent {\bf Hamilton's entropy. } It was noted in \cite{Frolov:1992xx} and \cite{Solodukhin:1994pc}  that
the 2d metric (\ref{metric2}), (\ref{gst}) that describes  a black hole in the string inspired
$\sigma$-model is also a solution to the field equations that follow from the
action
\be
N=\int_{\cal M} R\ln R~~,
\lb{Rln}
\ee
considered on a non-compact spacetime. This functional plays an important role
in the Hamilton's analysis of the Ricci flow on compact two-dimensional
manifold. It is a monotonic function along the flow on manifold with
positive curvature, $R>0$. A simple proof of this statement in the case of compact
manifold was suggested by Chow \cite{Chow}. Below we slightly modify this proof for
the case of non-compact asymptotically flat manifold.

Define a symmetric, trace free tensor  $M_{ij}=\nabla_i\nabla_j \psi
-{1\over 2}R g_{ij}$, where $\Delta \psi=R$, and a 1-form $X_i=\nabla_i
R+R\nabla_i\psi$. They are related as $X_i=2\nabla^j M_{ij}$. The direct
calculation shows that under the Ricci flow the functional (\ref{Rln}) evolves
as follows
\be
{dN\over d\lambda}=-2\int_{\cal M} |M_{ij}|^2-\int_{\cal M}{|X|^2\over R}
\leq 0~~.
\lb{dN}
\ee
A number of integrations by parts and the identity $\Delta\nabla_i f=\nabla_i
\Delta f+{1\over 2}R\nabla_i f$ are needed in order to get (\ref{dN}).
The right hand side of (\ref{dN}) vanishes when $M_{ij}=0$, and, hence,
$X_i=0$ (that gives relation $\psi=-\ln R$), that
is exactly the field equations that follow from (\ref{Rln}) by
variational principle.  This  is also condition for the steady solitonic
solution to the Ricci flow equation \cite{Hamilton2}.

Analyzing the convergence of the integrals let's focus on the class of static metrics
(\ref{metric2}) with $g(y)=1+O(1/y^k) $ with $k>0$ for large $y$. Then we have that
$\psi'_y=b_0+O(1/y^k)$ (as follows from (\ref{grad})), $R=O(1/y^{k+2})$ and, hence, $X_y=b_0R+O(1/y^{k+3})$.
It follows that for large $y$ we have that $R\ln R=O(y^{-k-2}\ln y)$,
$|X|^2/R\sim b_0^2R\sim O(1/y^{k+2}) $ and $|M_{ij}|^2\sim O(1/y^{2(k+1)})$.
Thus, all integrals in (\ref{Rln}) and (\ref{dN}) perfectly converge.

\section{The monotonicity of entanglement entropy}
\setcounter{equation}0
In order to show that the Ricci flow is irreversible in the mathematics
literature were introduced various definitions of entropy
\cite{Hamilton2}, \cite{Perelman} that have
the property to change monotonically under the flow.  On the other hand, if
the evolving metric  describes a black hole then this metric may have an intrinsic
gravitational entropy that measures the loss of information due to the
presence of the black hole horizon. So it is a natural question how this
entropy changes under the Ricci flow.  It should be noted that the
Bekenstein-Hawking entropy, that is standardly  associated with a black hole,
is defined ``on-shell'' for a metric that satisfies certain gravitational
equations obtained by variational principle from a gravitational action.
Entanglement entropy, discussed in section 2, is yet another entropy naturally defined for a black
hole. Its advantage is that it does not require knowledge of any particular gravitational
action and is defined ``off-shell'', i.e. for any metric that has properties
of black hole.

For the black hole geometry, at any given $\lambda$, entanglement entropy can be defined  by
first transforming metric to the Schwarzschild like form (\ref{metric2}) and then
applying the general formula (\ref{S}),
\be
S(\lambda)={c\over 12}\left(\int_0^{L_y}{dy\over g(y,\lambda)}(b_0-g'_y)+\ln
  g(L_y,\lambda)-2\ln (b_0\epsilon)\right)~~,
\lb{Sl}
\ee
where we take into account that $g'_y(0)=b_0$ for any $\lambda$.
$L_y$ is  the value of  $y$-coordinate at the boundary of the box.
We find that
\be
L_y=\int_0^L{g(x,\lambda)\over g_0(x)}dx~,~~\partial_\lambda L_y=(g'_y(L_y)-b_0)~~.
\lb{Ly}
\ee
In order to check whether the quantity (\ref{Sl}) is monotonic along
the Ricci flow we have to calculate its derivative with respect to $\lambda$
and look at the sign of this derivative. Before doing this it is convenient to
reshuffle a bit the terms in (\ref{Sl}).  Integrating by parts in
(\ref{Sl}) we get
\be
S(\lambda)={c\over 12}\left(b_0\int_\delta^{L_y}{dy\over g(y,\lambda)}+\ln g(\delta) -2\ln
  (b_0\epsilon)\right)~~,
\lb{SS}
\ee
where, for convenience, we introduced a small quantity $\delta$, (\ref{SS})
should be understood as limit $\delta\rightarrow 0$.
Calculating now the derivative of (\ref{SS}) with respect to $\lambda$,
using equations  (\ref{eq3}) and (\ref{Ly}) and integrating by parts we find
that
\be
\partial_\lambda S(\lambda)={c\over 12 g(\delta)} \left( \partial_\lambda
  g(\delta)+b_0(g'_y(\delta)-b_0) \right)~~.
\lb{SS1}
\ee
Using once again equation (\ref{eq3}) at $y=\delta$ we get that
\be
\partial_\lambda S(\lambda)={c\over 12} \left(
  g''_y(\delta)-{(g'_y(\delta)-b_0)^2\over g(\delta)}\right)~~.
\lb{SS2} \ee
The second term in (\ref{SS2}) is of order $\delta$
and vanishes in the limit $\delta\rightarrow 0$. We finally obtain
that \be
\partial_\lambda S(\lambda)=-{c\over 12}R(0,\lambda)~~,
\lb{SF}
\ee
where $R(0)=-g''_y(0)$ is the value of the Ricci scalar at horizon.

As we discussed this in section 3 the everywhere positive Ricci scalar
remains positive under the Ricci flow. Thus, if the initial black hole metric has
$R(x)>0$ for all $x>x_+$ then entanglement entropy is monotonically decreasing
under the Ricci flow. Note that the Ricci scalar is positive in the case of important
2d metrics: the Schwarzschild metric with $g_0(x)=1-{a\over
  x}$ and the string-inspired black hole metric with
$g_0(x)=1-ae^{-\Lambda x}$.  Since for the monotonicity of entanglement entropy
we need the Ricci scalar to be positive only at the horizon it is possible
that one can relax the condition on the Ricci scalar of the initial metric
and admit metrics with $R$ changing the sign far from the horizon. It would be
interesting to analyze this possibility.

It should be  noted that the evolution of the entanglement entropy does not
stop when the metric reaches the stationary point (\ref{gst}). Although the metric
components do not depend on $\lambda$ in this case, the size $L_y$ of the box in metric (\ref{metric2})
changes monotonically with $\lambda$, $\partial_\lambda
L_y=b_0(e^{-b_0L_y}-1)<0$. The monotonicity of the entropy for the stationary metric,
\be
\partial_\lambda S_{\tt st}=-{c\over 12}b_0^2<0~~,
\lb{Sl1}
\ee
then is entirely due to the monotonicity of the entropy as function of the size (\ref{monS}).

\section{Conclusions}
\setcounter{equation}0
With almost every known definition of entropy it is closely associated  a
property to change  monotonically under the change of certain parameters.
This property indicates the underlying irreversibility and motivates the actual usage of word
``entropy'' in each case. Entanglement entropy is an interesting quantity that
has various applications. Defined in  two dimensions for
the quantum mechanical non-gravitational systems it was shown to be monotonic
under the renormalization group flow \cite{Latorre:2004pk}. The gravitational
analog of the RG flow is the Ricci flow. In fact, the both flows are related
for  the two-dimensional $\sigma$-models whose target space is the Riemann
manifold. In particular, this manifold may be noncompact and may describe a black
hole as it was demonstrated in \cite{Witten:1991yr}.
The usual c-theorem does not immediately apply to the noncompact case
\cite{Polchinski:1987dy}. Thus, it is desirable to identify a quantity that
changes monotonically under the RG flow in this case.
In this note we have
shown that entanglement entropy, defined for  a two-dimensional black
hole metric, is monotonically decreasing  under the Ricci flow. The necessary condition for
the monotonicity is the positiveness of the Ricci scalar at the black hole
horizon.  In the case of the $\sigma$-model this allows us to assign with the
evolving  metric an important quantity that characterizes both the irreversibility
of the RG flow and the irreversible loss of information due to the black hole.

\bigskip

\bigskip

\noindent {\large \bf Acknowledgments}

\bigskip

\noindent
This work is supported in part by  DFG grant Schu 1250/3-1.


\begin{thebibliography}
\bibitem{Perelman}
   G. Perelman, ``The entropy formula for the Ricci flow and its geometric applications,"
math.DG/0211159.
\end{thebibliography}
\end{document}